\documentstyle[12pt]{article}
\input epsf.tex

\newcommand{\beq}{\begin{equation}}
\newcommand{\eeq}{\end{equation}}
\newcommand{\be}{\begin{equation}}
\newcommand{\ee}{\end{equation}}
\newcommand{\bea}{\begin{eqnarray}}
\newcommand{\eea}{\end{eqnarray}}
\newcommand{\req}[1]{~(\ref{#1})}

\newcommand{\gs}{\mbox{$g_s$}}            
\newcommand{\ap}{\mbox{$\alpha^\prime$}}  
\newcommand{\ls}{\mbox{$l_s$}}            
\newcommand{\g}[1]{g_{YM_{#1}}}

\def\href#1#2{#2}

\def\p{\partial}

\begin{document}

\baselineskip=15.5pt
\pagestyle{plain}
\setcounter{page}{1}

\begin{titlepage}
\begin{flushleft}
       \hfill                      {\tt hep-th/0801.4216}\\
       \hfill                       FIT HE - 08-01 \\
       \hfill                       KYUSHU-HET-111 \\
       \hfill                       
\end{flushleft}
\vspace*{3mm}

\begin{center}
{\huge Baryons with D5 Brane Vertex\\
\vspace*{2mm}
 and k-Quarks 
States}
\end{center}
\vspace{5mm}

\begin{center}
\vspace*{5mm}
\vspace*{12mm}
{\large Kazuo Ghoroku${}^{\dagger}$\footnote[1]{\tt gouroku@dontaku.fit.ac.jp},
Masafumi Ishihara${}^{\ddagger}$\footnote[2]{\tt masafumi@higgs.phys.kyushu-u.ac.jp},
%
}\\

\vspace*{2mm}

\vspace*{4mm}
{\large ${}^{\dagger}$Fukuoka Institute of Technology, Wajiro, 
Higashi-ku}\\
{\large Fukuoka 811-0295, Japan\\}
\vspace*{2mm}
{\large ${}^{\ddagger}$Department of Physics, Kyushu University, Hakozaki,
Higashi-ku}\\
{\large Fukuoka 812-8581, Japan\\
}

\end{center}

\begin{center}
{\large Abstract}
\end{center}
\noindent

We study baryons in $SU(N)$ gauge theories, 
according to the gauge/string correspondence based on IIB string theory.  
The D5 brane, in which $N$ fundamental strings are dissolved as a color singlet,
is introduced as the baryon vertex, and its configurations are studied.
We find point- and split-type of vertex. In the latter case, 
two cusps appears and they are connected by a flux composed of 
dissolved fundamental strings with a definite tension. 
In both cases, $N$ fundamental quarks are attached
on the cusp(s) of the vertex to cancel the surface term. 
In the confining phase, we find that
the quarks in the baryon feel the potential increasing linearly with 
the distance from the vertex. 
At finite temperature and in the deconfining phase, 
we find stable k-quarks "baryons", which are constructed
of arbitrary number of $k(<N)$ quarks. 

\vfill
\begin{flushleft}
January 2008
\end{flushleft}
\end{titlepage}
\newpage

\section{Introduction}
In the context of string/gauge theory correspondence 
\cite{jthroat,gkpads,wittenholo}, 
the large-$N$ dynamics of baryons can be related 
to the D5-branes embedded in $AdS$ space \cite{wittenbaryon, 
groguri}.
This point, in the context of the 
non-confining
${\mathcal N}=4$ supersymmetric Yang-Mills theory, have been studied 
in \cite{imamura,cgs}
using the Born-Infeld approach for constructing strings out of 
D-branes \cite{calmal,gibbons}. The fundamental strings
(F-string) are dissolved in the D5 brane in these approaches, and they
could flow out as separated strings from the singular point(s) 
appeared on the surface of the D5 brane. So we can consider the system
of the D5 brane and F-strings, which are out of the D5 brane, as baryon.
The approach in this direction has been applied to the 
confining gauge theories \cite{cgst,ima-04,ima-05}. 
In this case, the no force condition, a balance of the tensions of the D5 brane
and the F-strings, is imposed at their connected point(s) 
\cite{baryonsugra,imafirst}. However, in \cite{baryonsugra,imafirst},
the configuration of D5 branes with dissolved F-strings has not been considered.
In \cite{cgst}, the D5 configuration has been taken into account of, but 
the no force condition has been used only in a restricted direction, then
the structure of the F-strings is neglected.
However, taking into account of both structure of D5 brane and F-strings 
is important to obtain possible
baryon configurations as a system of the both objects.

Here, we study the plausible baryon configurations by embedding these
objects in two simple bulk backgrounds 
which correspond to a confining and a non-confining
$SU(N)$ gauge theories. In a confining theory, we can see
two kinds of tensions in the baryon. 
One is the string tension of the F-strings, which works on the
quarks in the baryon. And the
other is observed as the tension of the bundle composed of many F-strings.
The latter is found in a special configuration where the D5 brane is stretched. 

In the deconfining and high temperature phase, which is expressed by the AdS
Schwartzcshild background, we could find color non-singlet
baryon configuration constructed of the quarks with the number $k<N$. 
This configuration
could be generated due to the fact that $N-k$ F-strings of the baryon could disappear
into the horizon. As a result, the baryon is separated to free $N-k$ quarks and the remaining
k-quarks connected to the D5 baryon vertex as k-quarks baryon. 

The AdS$_5$ background expresses also the deconfining phase, so 
the k-string state has been 
found also in this theory with a constraint for the number $k$ as
$5N/8<k<N$ in \cite{baryonsugra,imafirst}. In these approaches,
any configuration of D5 vertex 
has not been considered as a solution of the brane action. In our case, 
however, such structures are taken into account of, then the lower
bound for $k$ is removed as a result. Due to
the geometrical freedom of the brane and the
F-strings, any $k$ quarks baryon is possible.

In Section \ref{eqnsec} we give our model and D5-brane action
with non-trivial $U(1)$ gauge field which represents the dissolved F-strings.
And the equations of motion for D5 branes are given. 
In section
\ref{point3dsec}, we give
a point vertex solution and study possible baryon configurations
in the confining phase. We show the linear relation of the baryon mass
and the distance of a quark in the baryon from the vertex.
In section \ref{split3dsec}, 
the split vertex solutions are studied. This solution is constructed by the
bundle of F-strings flux with a definite tension between two cusps
on ${\bf S}^5$. And we estimate the tension of this flux. 
In section 5, we study the baryon in the deconfinement phase at 
finite temperature, and stable color non-singlet baryons are shown. And 
in a final section, we summarize our results and discuss future directions.


\section{Set of the Model}
\label{eqnsec}

We derive the equations for a D5-brane embedded in the appropriate
10d background.
As a specific supergravity background, we consider the following 10d
background in string frame given by a non-trivial dilaton $\Phi$ and
axion $\chi$ \cite{Liu:1999fc,GY},
\beq
ds^2_{10}= e^{\Phi/2}
\left(
\frac{r^2}{R^2}A^2(r)\eta_{\mu\nu}dx^\mu dx^\nu +
\frac{R^2}{r^2} dr^2+R^2 d\Omega_5^2 \right) \ .
\label{non-SUSY-sol}
\eeq
First, we consider 
the supersymmetric solution
\beq 
A=1, \quad e^\Phi= 
1+\frac{q}{r^4} \ , \quad \chi=-e^{-\Phi}+\chi_0 \ ,
\label{dilaton} 
\eeq
with self-dual Ramond-Ramond field strength
\begin{equation}\label{fiveform}
G_{(5)}\equiv dC_{(4)}=4R^4\left(\mbox{vol}(S^5) d\theta_1\wedge\ldots\wedge d\theta_5
-{r^3\over R^8}  dt\wedge \ldots\wedge dx_3\wedge dr\right),
\end{equation}
where $\mbox{vol}(S^5)\equiv\sin^4\theta_1
\mbox{vol}(S^4)\equiv\sin^4\theta_1\sin^3\theta_2\sin^2\theta_3\sin\theta_4$.
The 
solution (\ref{dilaton}) is dual to $N=2$ super Yang-Mills theory with
gauge condensate $q$ and is chirally symmetric. 


\vspace{.5cm}
Here the baryon is constructed from the vertex and $N$ fundamental strings.
In the string theory, the vertex is considered as
the D5 brane wrapped on an
${\bf S}^{5}$ on which $N$ fundamental strings terminate and they are dissolved
in it \cite{wittenbaryon,groguri}. 
Then the D5-brane action is written as by the Born-Infeld plus
Chern-Simons term
\begin{eqnarray}\label{d5action}
S_{D5}&=&-T_{5}\int d^6\xi
 e^{-\Phi}\sqrt{-\det\left(g_{ab}+2\pi\ap F_{ab}\right)}+T_{5}\int
\left(2\pi\ap F_{(2)}\wedge c_{(4)}\right)_{0\ldots 5}~,\\
g_{ab}&\equiv&\p_a X^{\mu}\p_b X^{\nu}G_{\mu\nu}~, \qquad
c_{a_1\ldots
a_4}\,\equiv\,\p_{a_1}X^{\mu_1}\ldots\p_{a_4}X^{\mu_4}C_{\mu_1\ldots\mu_4}~.\nonumber
\end{eqnarray}
where $T_5=1/(\gs(2\pi)^{5}\ls^{6})$ is the brane tension.
The Born-Infeld term involves
the induced metric $g$ and the $U(1)$ worldvolume
field strength $F_{(2)}=d A_{(1)}$.
The second term is the Wess-Zumino coupling of the
worldvolume gauge field, and it is also written as
$$
S = -T_5 \int d^6\xi~ e^{-\Phi}
     \sqrt{-\det(g+F)} +T_5 \int A_{(1)}\wedge G_{(5)}~,
$$
in terms of (the pullback of)
the background five-form field strength $G_{(5)}$, which effectively
endows the fivebrane with a $U(1)$ charge proportional to the
${\bf S}^{5}$ solid angle that it spans.
Namely, 
\beq\label{quantized flux}
\int_{S^4}d^{4}\theta {\partial L\over \partial F_{(2)}}=
\int_{S^4}d^{4}\theta {\partial L\over \partial B_{(2)}}=
 k T_F
\eeq
where $T_F=1/(2\pi\alpha')$ and $k\subset Z$.

\vspace{.5cm}
Then we give the embedded configuration of the D5 brane in the given 10d
background. At first, we fix its world volume as
$\xi^{a}=(t,\theta,\theta_2,\ldots,\theta_5)$.
For simplicity we
restrict our attention to $SO(5)$ symmetric configurations of the 
form
$r(\theta)$, $x(\theta)$, and $A_t(\theta)$ (with all other fields 
set to
zero), where $\theta$ is the polar angle in spherical coordinates.
The action then simplifies to
\be \label{d3action}
S= T_5 \Omega_{4}R^4\int dt\,d\theta \sin^4\theta \{ -
  \sqrt{e^{\Phi}\left(r^2+r^{\prime 2}+(r/R)^{4}x^{\prime 2}\right)
   -F_{\theta t}^2}  +4 A_t \},
\ee
where $\Omega_{4}=8\pi^{2}/3$ is the volume of the unit four-sphere.

The gauge field equation of motion following from this action reads
$$
\partial_\theta D = -4 \sin^4\theta,
$$
where the dimensionless displacement 
is defined as the variation of the
action with respect to $E=F_{t\theta}$, namely
$D=\delta \tilde{S}/\delta F_{t\theta}$ and 
$\tilde{S}=S/T_5 \Omega_{4}R^4$. The solution to this equation 
is
\be \label{d}
D\equiv D(\nu,\theta) = \left[{3\over 2}(\nu\pi-\theta)
  +{3\over 2}\sin\theta\cos\theta+\sin^{3}\theta\cos\theta\right].
\ee
Here, the integration constant $\nu$ is expressed as $0\leq\nu=k/N\leq 1$,
where $k$ denotes the number of Born-Infeld strings emerging from one of the pole of
the ${\bf S}^{5}$. Next, it is convenient to eliminate the gauge 
field
in favor of $D$ and Legendre transform the original Lagrangian to
obtain an energy
functional of the embedding coordinate only:
\be \label{u}
U = {N\over 3\pi^2\alpha'}\int d\theta~e^{\Phi/2}
\sqrt{r^2+r^{\prime 2} +(r/R)^{4}x^{\prime 2}}\,
\sqrt{V_{\nu}(\theta)}~.
\ee
\be\label{PotentialV}
V_{\nu}(\theta)=D(\nu,\theta)^2+\sin^8\theta
\ee
where we used $T_5 \Omega_{4}R^4=N/(3\pi^2\alpha')$.
Using this expression (\ref{u}), we solve the D5 brane configuration
in the followings.

\section{The Point Vertex}\label{point3dsec}

In this section we study solutions which correspond to a baryon
localized at a particular point in our 4d space-time. To localize the vertex 
in $x$, we set as $x'=0$, and
the equation of motion for $r(\theta)$ is obtained from (\ref{u}) as
\beq
\partial_{\theta}\left({r^{\prime}\over\sqrt{r^2+(r^{\prime})^2}}\,
\sqrt{V_{\nu}(\theta)}\right)
-\left(1+{r\over 2}\partial_r\Phi\right){r\over\sqrt{r^2+(r^{\prime})^2}}\,
\sqrt{V_{\nu}(\theta)}=0~.
\eeq
For our background solution (\ref{dilaton}), it is rewritten as
\beq\label{eq-r}
\partial_{\theta}\left({r^{\prime}\over\sqrt{r^2+(r^{\prime})^2}}\,
\sqrt{V_{\nu}(\theta)}\right)
-{1-q/{r^4}\over 1+q/{r^4}}{r\over\sqrt{r^2+(r^{\prime})^2}}\,
\sqrt{V_{\nu}(\theta)}=0~.
\eeq


In the case of $q=0$, the background reduces to $AdS_5\times S^5$, where
the equation is analytically solved and well studied in \cite{cgs}. 
In this case, however, the quarks are not confined. In order to see the
baryon configuration, we should consider the case of non-BPS confinement
phase, which is realized in our model for non-zero $q>0$.
The parameter $q$ represents the gauge condensate in the 4d
YM theory, and provide the tension of the meson states. So we could expect 
to be able 
to extract the characteristic properties of the baryon in the confinement phase. 

\begin{figure}[htb]
\centerline{{\epsfxsize=7cm
\epsfbox{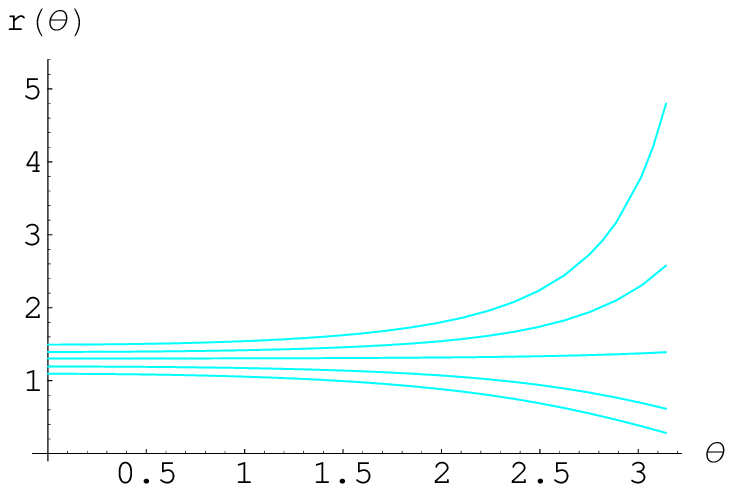}}{\epsfxsize=7cm
\epsfbox{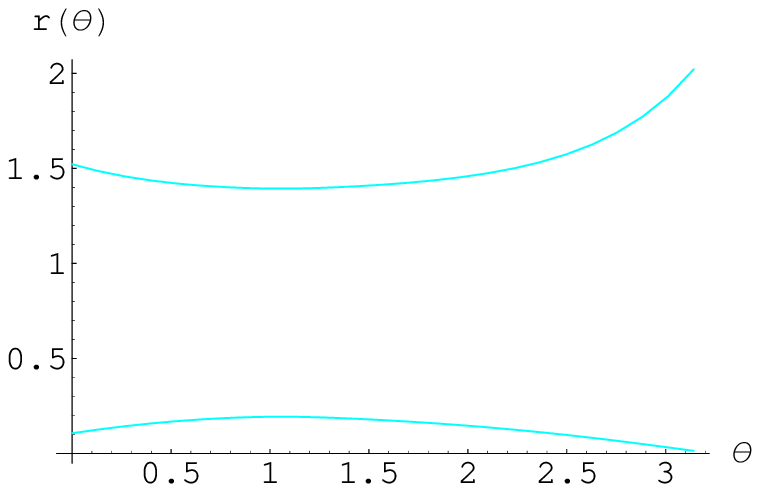}}}
\caption{\small Family of solutions for $q=2.8$ and $r(0)=q^{1/4}+\epsilon$ with various $\epsilon$, $\epsilon=0.2,~0.1,~.01, -0.2, -0.3$ for the curves
from the above. The right hand side shows the solutions for $\nu=0.2$
and $r(\theta_c)=q^{1/4}+ 0.1$ (the upper) and $q^{1/4}- 1.1$ (the lower). }
\label{D3tubes}
\end{figure}

However, for $q\neq 0$, it is difficult to find an analytical solution, so
we try to solve the above 
equation numerically. The "potential" $V(\theta)$ has three extremum points.
In solving the equation, we impose the boundary condition at one of these
point, $\theta=\theta_c$, which is the minimum of $V_{\nu}(\theta)$ 
and is given by the solution of
\be \label{nu}
\pi\nu=\theta_{c}-\sin\theta_{c}\cos\theta_{c}~.
\ee
Then the boundary conditions are set as 
\beq
 r(\theta_c)=r_0, \quad {\partial r(\theta_c)\over \partial \theta}=0~,
\eeq
where $r_0$ is a parameter which determine the configuration of D5 brane.
The typical solutions are shown in the Fig. \ref{D3tubes}. These solutions
have cusps at the pole points, $\theta=\pi$ and $\theta=0$ for $\nu=0.2$ and
only at $\theta=\pi$ for $\nu=0$.

\vspace{.5cm}
For simplicity, we consider here the case of $\nu=0$.
In this case, $\theta_c=0$ and 
the solution $r(\theta)$ is smooth at $\theta=0$ but it has a
cusp at $\theta=\pi$ on $S^5$.
The solutions depend on
$q$ and $r_0$ through the ratio $\zeta=r_{0}/q^{1/4}$, and they
are classified to two types, (A) and (B),
by $\zeta$ as follows;

\vspace{.3cm}
\noindent {\bf (A) $\zeta\geq 1$;} 

(i) For $\zeta>>1$, the solution is near the BPS `tube' solution \cite{cgs}, 
which arrives at $\theta=\pi$ only at $r=\infty$. 

(ii) When $\zeta$ decreases toward $\zeta=1$, the solution begins to
tilt and crosses the symmetry axis,
$\theta=\pi$, at a finite value of $r=r_c$. And $r'(\pi)$ is 
finite then it forms a cusp.

(iii) And we notice that 
$r(\theta)=q^{1/4}$ at $\zeta=1$, and this constant value of $r$ 
is an exact solution to the equations of motion.
Its shape is completely spherical. This is the only one allowable
constant solution. 

For this solution, we obtain from (\ref{u}) the following
\be\label{junction-e0}
U ={N\over 3\pi^2\alpha'}{2^{1/2}q^{1/4}}j(0)~,
\ee
where $$j(\nu)=\int_0^{\pi}\sqrt{V_{\nu}(\theta)}$$ and it is
estimated as $j(0)=10.67$ for $\nu=0$. This
represents the vertex energy for the constant solution of $r$. A similar
solution has been given in \cite{baryonsugra,imafirst} for AdS${}_5$ background.
In our case, however, this solution corresponds to a special one in our 
non-conformal background. Actually, this solution disappears into the horizon
in the AdS${}_5$ limit of $q\to 0$. 

\vspace{.3cm}
\noindent {\bf (B) $0<\zeta<1$ ;}
In this case, the term $1-q/r^4$ in the second term
of Eq. (\ref{eq-r}) turns to negative then $r$ decreases with $\theta$.
This implies that the north and south 
poles on $S^5$ 
are opposite to the one of the solution for $\zeta>1$, 
then the direction of the force coming
from the D5 tension at the cusp is also reversed. 
Further,
the minimum, $r(\pi)$, approaches to zero for $r_0\to 0$, but 
the brane could not arrive at 
$r=0$ since the action diverges there
even if how small $r_0$ is. In other words, an infinite
energy is necessary to realize $r(\pi)=0$. In other words, the D5 vertex is never
absorbed in the horizon due to the confinement force, and $\zeta$ is restricted as
$\zeta>0$.

\vspace{.3cm}
\subsection{Baryon configuration and no-force condition}
\begin{figure}[htb]
\centerline{{\epsfxsize=14cm
\epsfbox{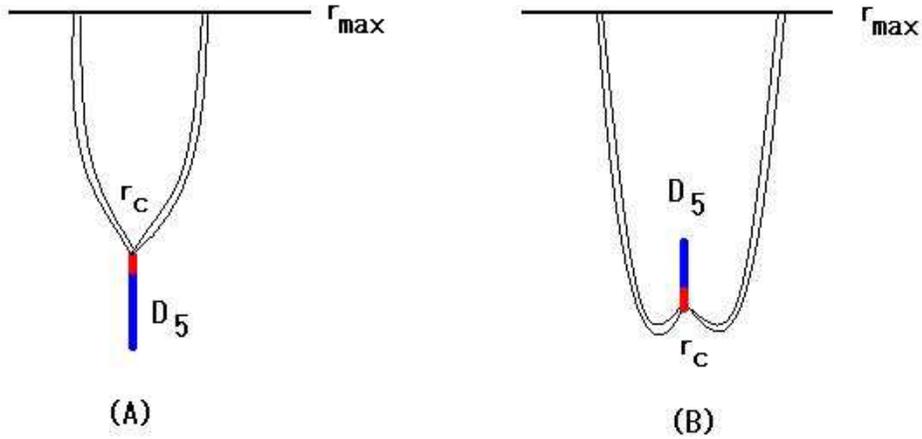}}}
\caption{\small Point vertex baryon. For the solutions of (A) $r_c>r_0$ 
and (B) $r_c<r_0$.}
\label{baryon-pic}
\end{figure}

The schematic configurations of the baryons formed of the two types of
vertex (A) and (B) are shown in the Fig. \ref{baryon-pic}.
In the present case, 
the fundamental strings, which are dissolved in the D5 brane, should
flow out from the cusp at $\theta=\pi$.
The D5 brane configuration is singular at this point. This singularity is cancelled out
by the boundary term of the fundamental string stretching from this point.
This cancellation is equivalent to the so called no force condition, 
namely the cancellation
of the tension forces among the fundamental strings and the D5 brane at the 
cusp point. It is possible to consider various configurations 
of D5 brane and strings which satisfy the no force condition. They are supposed
as the various possible flowing of the fundamental strings from the D5 brane.

Then the fundamental strings coming out from the D5 brane
could stretch separately
to any direction when they are allowed dynamically. The dynamical conditions 
to be considered are
separated to two parts, (a) the equations of motion of F-strings 
in the given background and (b) 
the no force conditions at the cusp point.

As for the condition (b), when the stretching direction
of the F-strings is restricted to 
the $r$ direction only as considered in \cite{cgst}, then the
resultant baryon configuration is reduced to the BPS flux tube of D5
since the tension force of the F-strings is stronger than the one of the D5 vertex.
However we could consider other configurations
of F-strings which spread 
also to the $x$ direction. As a result, the force in the $r$-direction is weakened
and the configurations given in the Fig. \ref{baryon-pic} become possible.

Such a string configuration has been considered in \cite{baryonsugra}, but the authors
in \cite{baryonsugra} did not take into account of
the structure of the D5 brane except for the constant $r$ 
configuration as mentioned above. In the present model, more general
configurations than the one considered in \cite{baryonsugra} are studied.

\vspace{.3cm}
Here we suppose that the two end points of F-string are connected to
the cusp $r=r_c$ of the D5 vertex and the other probe brane, 
for example the D7 probe brane,
which corresponds to the brane providing the flavor quarks and 
is put at an appropriate position,
$r=r_{\rm max}(>r_c)$. Here $r_{max}$ plays a role of the cut off of $r$,
and we do not give a D7 brane configuration with the attached strings.
This problem is postponed to the future works. 
The quark mass in this case is given approximately
by $T_F (r_{\rm max}-r_c)$ for the AdS$_5$ background.

\vspace{.3cm}
Now we turn to the no force condition at $r=r_c$.
In the $r$ direction, the tensions of the fundamental strings and
the one of the D5 brane should make a balance. 
As for the $x$ direction, the balance should be realized by F-strings
themselves.
For D5 brane, the tension in the $r$ direction is estimated in terms of $U$
given by \req{u} under the variation $r_{c}\to
r_{c}+\delta r_{c}$\cite{cgst}, it can be seen from \req{u}, by using 
the Euler-Lagrange equation,
that the energy of the brane changes only by a surface term at $r_c$,
\be \label{varcalc}
\frac{\partial U}{\partial r_{c}}
 =NT_F e^{\Phi/2}\frac{r_{c}^{\prime}}{\sqrt{r_{c}^{\prime 2}+r_{c}^2}}~,
\ee
where $r_{c}^{\prime}=\partial_{\theta}r|_{\theta=\pi}$.
From this equation, we can see that the direction of the force is reversed
at the point $\zeta=1$ when $\zeta$ changes from (A) $\zeta>1$ to (B) $\zeta<1$
since the sign of $r_{c}^{\prime}$ changes.
This behavior is important in our model as found below.

\vspace{.3cm}
As for the fundamental string, which extends to
the $x$-direction, its action is written as
\be \label{Fstring}
S_F = -{1\over 2\pi\alpha'}\int dtdx~e^{\Phi/2}
\sqrt{r_x^2+ (r/R)^{4}}~,
\ee
where $r_x=\partial r/\partial x$ and the world sheet coordinates are set as $(t,x)$. 
Then, its energy and $r$-directed tension at the point $r=r_c$ are obtained as follows
\be \label{FstringU}
U_{\rm F} = T_F\int dx~e^{\Phi/2}
\sqrt{r_x^2+ (r/R)^{4}}~,
\ee
\be \label{FstringdU}
\frac{\partial U_F}{\partial r_{c}}
 =T_F e^{\Phi/2}\frac{r_{x}}{\sqrt{r_{x}^{2}+(r_{c}/R)^4}}~,
\ee
For the remaining N-1 strings,
it is possible to take their world sheet coordinates in a different
directions from the $x$ direction.
Here, for the simplicity, we assume that they extend in the same plane,
for example the $x$-$y$ plane,
and their forces make a valence with the same tension in this plane. 
The quarks are put on a circle whose center is the D5 vertex.
Then,
the forces of the strings in the $r$-direction are added up, and
its value is N times of (\ref{FstringdU}) and it should keep a valence
with the one of the D5 brane. 
We notice that the direction of this tension force is also reversed
at $\zeta=1$.

Thus we find the following no force 
condition as discussed in
\cite{baryonsugra,imafirst} with a slightly different setting,
\be \label{no-force}
r_{x}^{(c)}=r_c'{r_c\over R^2}~,
\ee
at $r=r_c$, where $r_{x}^{(c)}$ denotes the value of $r_{x}$ at $r=r_c$. 
Here the sign of $r_{x}^{(c)}$ and $r'_c$ should be the same. 

\subsection{Baryon mass and distance between quark and vertex}
Thus we find the baryon configuration as depicted schematically
in the Fig.\ref{baryon-pic}. For small baryon mass, its configuration
is expressed by the figure (A), and the configuration changes to (B) when
the mass increases. In the configuration (B), the fundamental string could become
long with increasing baryon mass.
As a result, any large baryon mass is realized by the configuration
(B). 

The energy (or the mass)
of the baryon is therefore given as follows,
\beq
E=NE_F+E_J^{(S)}
\eeq
where $E_F$ is the F-string part and
$E_J^{(S)}$ represents the D5 brane part, i.e. the baryon vertex. 
The latter is obtained from
(\ref{u}) by setting as $x'= 0$, 
\beq\label{pvertex-e}
E_J^{(S)}= {N\over 3\pi^2\alpha'}\int_0^{\pi} d\theta~e^{\Phi/2}
({r^2+r^{\prime 2}})^{1/2}\,
\sqrt{V_{\nu=0}(\theta)}~.
\eeq
While the baryon vertex is seen as a point in our 4d space-time, it
has a structure in inner space and has a
finite value of $E_J^{(S)}$. 

Although $r_0$ is not equal to $q^{1/4}$ in general,
the rough estimation of this energy
is given analytically in this limit of $r_0=q^{1/4}$ or $\zeta=1$.
The D5 brane is squashed, at this point,
to a point in the $r$-direction. Then we obtain
\be\label{junction-e}
E_J^{(S)}\vert_{r_0=q^{1/4}}={N\over 3\pi^2\alpha'}{2^{1/2}q^{1/4}}j(0)~,
\ee
where 
$j(0)=10.67$ as given above.
Since $q$ is written as 
$q=\lambda\langle F_{\mu\nu}^2\rangle\alpha'~^4$ \cite{Liu:1999fc}, 
this vertex energy 
increases with the 'tHooft coupling like $\lambda^{1/4}$
when $\langle F_{\mu\nu}^2\rangle$ is fixed. 
On the other hand, we find that the F-string tension
is independent of $\lambda$ as seen below. 
Then, the vertex energy is expected to become
large and the main part of the baryon mass at large 'tHooft coupling.
 
\vspace{.5cm}
Next, we turn to the energy of the
F-string part. From (\ref{Fstring}), we can set the following
constant $h$,
\be \label{Fstringh}
e^{\Phi/2}{r^4\over R^4\sqrt{r_x^2+ (r/R)^{4}}}=h~.
\ee
Then, by eliminating $r_x$ in terms of the above equation with a constant $h$,
we get the formula of the distance $L$ between the quark and the vertex
and the string energy $E$ as 
\be\label{F-length}
L_{q-v}=R^2\int_{r_{c}}^{r_{\rm max}}dr{1\over r^2\sqrt{e^{\Phi}r^4/(h^2R^4)-1}}
\ee
\be\label{F-energy}
E_F=T_F\int_{r_{c}}^{r_{\rm max}}dr{e^{\Phi/2}\over \sqrt{1-h^2R^4/(e^{\Phi}r^4)}}
\ee
These formula are equivalent to the one of the mesons made of quark
and anti-quark except for 
the lower bound of the r-integration,
which is given here as the D5 cusp point $r_c$. 
And the no force condition (\ref{no-force}) is imposed
at this point.

\vspace{.3cm}
The equations (\ref{F-length}) and (\ref{F-energy}) 
are evaluated for the solutions (A) and (B)
separately since the string shapes are different in the two cases.
First, we consider the solutions of (B). In this case, $r_x$ is negative at 
$r_c$ then $r$ decreases with 
increasing $|x-x(r_c)|$ after it departed from the cusp point
$x(r_c)$, 
but the string can not never reach at the
horizon $r=0$ since the action diverges at this point. Thus $r$ reaches at the minimum
$r(\equiv r_{\rm min})$ at some $x=x_0$, where $r_x=0$, 
then it begins to increase toward $r_{\rm max}$ (see Fig. \ref{baryon-pic}).
The shape of this F-string is determined as the extremum of $U_F$, 
namely it is given 
as a solution of the equation of motion derived from
(\ref{FstringU}). The total configuration of the baryon made of
F-strings and the D5 vertex is controled by the one parameter $r_0$. For a given
$r_0$, all the value of $r'(\pi)$, $r_c$, $r_{\rm min}$ and $r_x\vert _{r_c}$ 
are determined, so the energy of
the baryon is also determined.

Therefore,
when a value of $r_0(<q^{1/4})$ is fixed, it is convenient to separate
the F-string to (i) the region of 
$r_{\rm min}<r<r_{\rm max}$ and of (ii) $r_{\rm min}<r<r_{c}$. 
Then the energy can be written as
\beq
E_F^{(B)}=E_F^{(B_i)}+E_F^{(B_{ii})}~.
\eeq
The first term
corresponds to the half
of the meson configuration made of quark and anti-quark.
When the energy becomes large or F-string grows long,  
$E_F^{(B_i)}$ is approximated by the
following formula \cite{GY}, 
$$ E_F^{(B_i)}=m_q^{\rm eff}+\tau_{M}{L_{q-\bar{q}}\over 2}$$
where
\be \label{f-tension}
\tau_M=T_F\sqrt{q\over R^4}~.
\ee
\be\label{m-eff}
m_q^{\rm eff}=T_F\int_{r_{\rm min}}^{r_{\rm max}}dr e^{\Phi/2}
\ee
In this calculation, 
the constant $h$ is taken as
\be
h=e^{\Phi(r_{\rm min})/2}\left({r_{\rm min}\over R}\right)^2
  =e^{\Phi(r_c)/2}\left({r_c\over R}\right)^2{1\over\sqrt{(r'(r_c)/r_c)^2+1}}~.
\ee
Due to this boundary condition,
$r_{\rm min}$ is determined by using 
$r_c$, then by $r_0$ since $r_c$ is determined by $r_0$ as mentioned
above.
The remaining part of the F-string is obtained as
\beq
E_F^{(B_{ii})}=T_F\int^{r_{c}}_{r_{\rm min}}dr{e^{\Phi/2}\over \sqrt{1-h^2R^4/(e^{\Phi}r^4)}}~.
\eeq

\vspace{.3cm}
Next, we turn to 
the solution (A), whose configuration
is shown in the Fig. \ref{baryon-pic} (A). In this case, since $r'(\pi)>0$,
then the $r$ of the F-string increases from $r_c$ towards $r_{\rm max}$
monotonically. There is no point of $r_x=0$. So the configuration of F-string
is obtained in terms of 
(\ref{F-length}) and (\ref{F-energy}) by the setting of the lower bound of
$r$ integration as $r_c$ and with the boundary condition at this point,
\be
h=e^{\Phi(r_c)/2}\left({r_c\over R}\right)^2{1\over\sqrt{(r'(r_c)/r_c)^2+1}}~.
\ee
In this case, the energy of the F-string is given as
\beq
E_F^{(A)}=T_F\int_{r_{c}}^{r_{\rm max}}dr{e^{\Phi/2}\over \sqrt{1-h^2R^4/(e^{\Phi}r^4)}}~.
\eeq 
The above $h$ could take its minimum, $e^{\Phi(r_c)/2}\left({r_c\over R}\right)^2$,
at $r'(r_c)=0$, which is realized at $r_0=q^{1/4}$, and the distance $L_{q-v}$ becomes the
maximum in the configuration (A). Actually, we can estimate the maximum
of $L_{q-v}$ as,
\be\label{Lmax}
L_{q-v}^{\rm max}=R^2\left({4\pi^2\over q}\right)^{1/4}{\Gamma(3/4)\over \Gamma(1/4)}
\ee
where we take $r_{\rm max}=\infty$. So the baryon is expressed by the solution (A)
for $0<L_{q-v}\leq L_{q-v}^{\rm max}$ and by the (B) for $L_{q-v}^{\rm max}<L_{q-v}$.

\begin{figure}[htb]
\centerline{
{\epsfxsize=10cm
\epsfbox{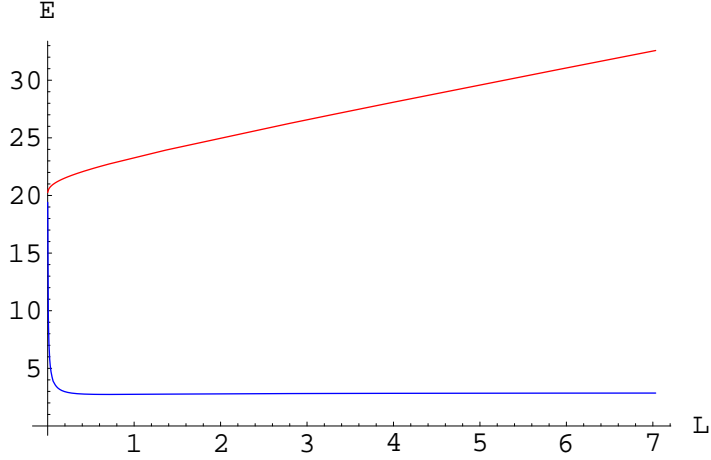}}}
\caption{\small $E$-$L_{q-v}$ plots for $q=2.0$, $R=1$, $r_{\rm max}=20$.
The upper (lower) curve shows the baryon energy $E$ (the vertex 
energy $E_{\rm D5}$).} 
\label{D5baryon}
\end{figure}

\vspace{.3cm}
Now we can calculate the total energy of the baryon $E$ versus $L_{q-v}$. 
The energy is given as
\beq
E=N E_F+E_{\rm D5}
\eeq
where $E_F=E_F^{(A)}$ or $E_F^{(B)}$ for short 
or long $L_{q-v}$ respectively. Assuming that all fundamental quarks 
are put at the same distance from the vertex, $E$ is numerically estimated 
as a function of $L_{q-v}$. An example is shown in
the Fig. \ref{D5baryon} for $r_{\rm max}=20$. 

In this figure, the value of $E$ at $L=0$ shows the value of $E_J^{(S)}$
with $r_c=r_{\rm max}$, namely the pure D5 brane energy. When $L$ begins
to increase, $E_J^{(S)}$ decreases rapidly with $L$ and approaches to a
constant, and $E_F$ becomes dominant. It is expressed by $E_F=E_F^{(A)}$
for small $L$, in the region of $L\leq 0.7$, and by $E_F=E_F^{(B)}$ for
$L\geq 0.7$ in the present case. 
We can see for large $L$ that $E$ increases with $L_{q-v}$
linearly with the tension,
\beq
 N\tau_M/2\, ,
\eeq
where $\tau_M$ is given in (\ref{f-tension}). This tension is equal to the
sum of the one of the independent F-strings.

\section{Split Vertices}
\label{split3dsec}

\vspace{.5cm}
The equations of motion for D5 embedding are solved here by adding the freedom
of $x(\theta)$ without the restriction $x'=0$. In this case, it is
convenient to use a parametric Hamiltonian formalism as in \cite{cgst}.
 
First we rewrite the energy (\ref{u}) in terms of a 
general
worldvolume parameter $s$ defined by
functions $\theta=\theta(s)$, $r=r(s)$, $x=x(s)$ as:
\be \label{upar}
U = {N\over 3\pi^2\alpha'}\int ds~e^{\Phi/2}
\sqrt{ r^2\dot{\theta}^2 + \dot{r}^2+(r/R)^{4}\dot{x}^2}~
\sqrt{V_{\nu}(\theta)},
\ee
where dots denote derivatives with respect to $s$.
The momenta conjugate to $r$, $\theta$ and $x$ are
\be \label{mom}
p_r=\dot{r}\Delta, \quad
p_{\theta}=r^2\dot{\theta}\Delta, \quad
p_{x}=(r/R)^{4}\dot{x}\Delta, \quad
\Delta=e^{\Phi/2}\frac{\sqrt{V_{\nu}(\theta)}}
   {\sqrt{ r^2\dot{\theta}^2 + \dot{r}^2+(r/R)^{4}\dot{x}^2}}~.
\ee
Since the Hamiltonian that follows from the action\req{upar} vanishes 
identically
due to reparametrization invariance in $s$. 
Then we consider the following identity
\be \label{ham}
2\tilde{H} =p_r^2 + \frac{p_{\theta}^2 }{r^2}+\frac{R^4}{r^4}p_{x}^2-
\left(V_{\nu}(\theta) \right) e^{\Phi}=0~.
\ee
This constraint can be taken as the new Hamiltonian
in order to get the following canonical equations of motion, 
\bea\label{Heom}
\dot r &=& p_r~,~~ \dot p_r =\frac{2}{r^5}p_x^2 R^4
     + \frac{p_\theta^2}{r^3}+{1\over 2}
        \left(V_{\nu}(\theta)\right)~ e^{\Phi}\partial_r\Phi,\\
\dot\theta &=&\frac{p_\theta}{r^2}~, ~~\dot p_\theta = -6\sin^4\theta \left(\pi\nu-\theta+ 
\sin\theta\cos\theta
\right)~e^{\Phi}, \\
\dot x &=& \frac{R^{4}}{r^{4}}p_{x},~~\dot{p}_{x}= 0
\eea
In solving these equations, the initial conditions should be
chosen such that $\tilde H=0$.

\begin{figure}[htb]
\centerline{{\epsfxsize=6cm\epsfbox{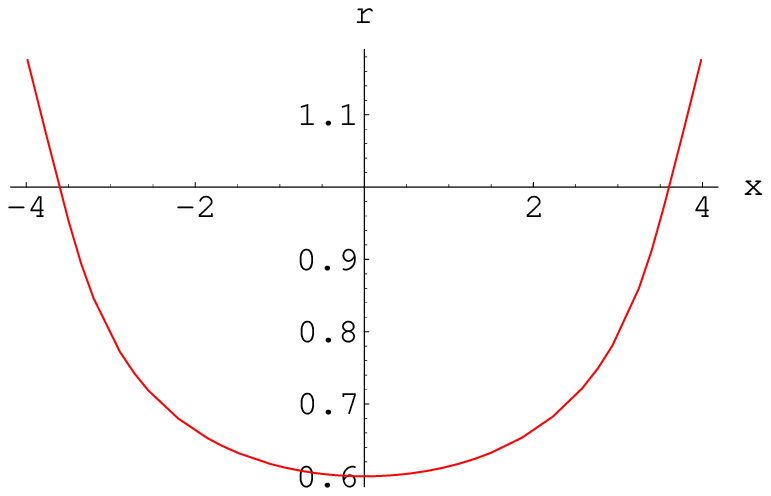}}
{\epsfxsize=6cm\epsfbox{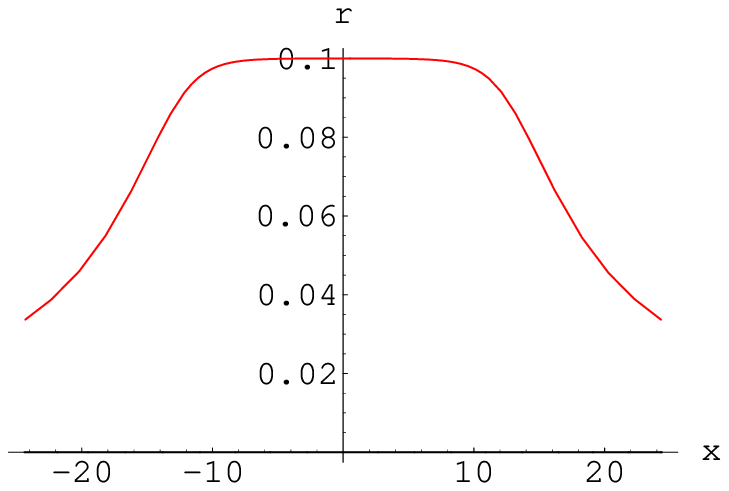}}
}
\begin{picture}(0,0)
\end{picture}
\caption{\small The solution with cusps
for $R=1$, $q=2$ and $\nu=0.5$. The left figure is for $r(\theta=\pi/2)=0.6$ and $r(\pi)=r(0)=0.490507$. The right one is for $r(\pi/2)=0.1$ and 
$r(\pi)=r(0)=0.0336822$. }
\label{split-rx-3}
\end{figure}
\vspace{.3cm}
We now solve these equations
numerically to obtain a configuration spreading 
in the $x$-direction with $p_{x}\neq 0$
(i.e. $x^{\prime}\neq 0$). We find two types of configurations of
D5 brane, the U shaped and the cap ($\bigcap$) shaped one.
Their example solutions are shown in the Fig. \ref{split-rx-3}.
The U shaped one is obtained for large $r(0)$ or small mass baryon,
and the cup shaped is obtained when $r(0)$ or baryon mass increses.

\begin{figure}[htb]
\centerline{{\epsfxsize=14cm
\epsfbox{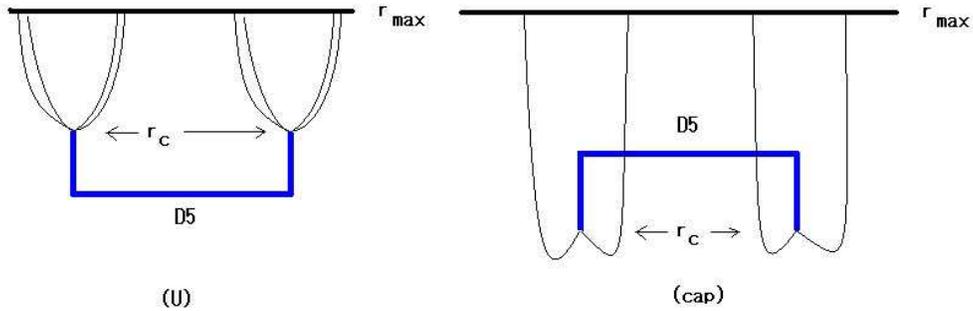}}}
\caption{\small Split vertex baryon. For the solutions of U shaped $r_c>r_0$ 
and cap shaped $r_c<r_0$.}
\label{baryon-p}
\end{figure}

Since $r$ and $\partial r/\partial\theta$ are finite at the end points of 
these configurations, the end points of both sides are the cusps. Then
the baryon configurations given here are
splited into two distinct cusps, which are connected to $\nu N$ and
$(1-\nu)N$ quarks respectively for the case of a given value of $\nu$.
We can see that the cusps of U shaped configuration are the type 
(A) which is given in the previous section, and type (B) cusps are seen
for the cap shaped one. Then the quarks are attached as depicted in the 
Fig. \ref{baryon-p} by considering the no force condition.  

\begin{figure}[htb]
\centerline{{\epsfxsize=5cm
\epsfbox{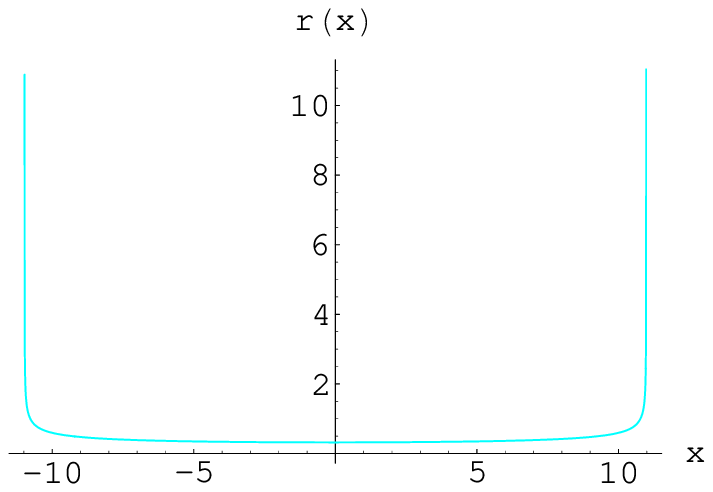}}
{\epsfxsize=5cm\epsfbox{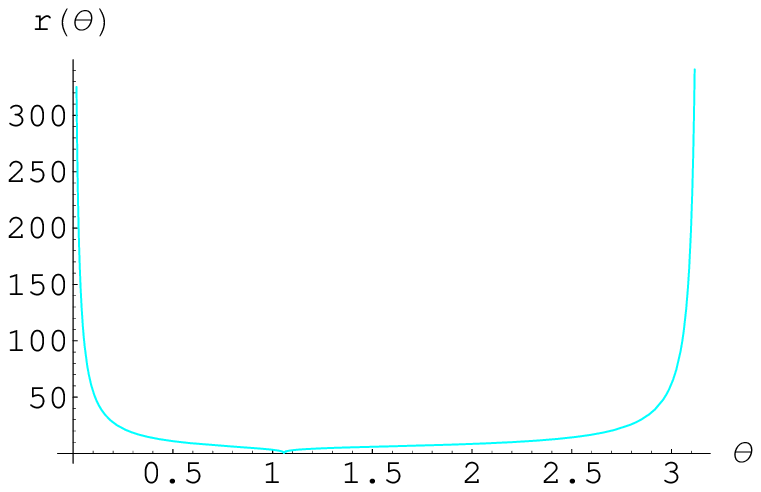}}{\epsfxsize=5cm\epsfbox{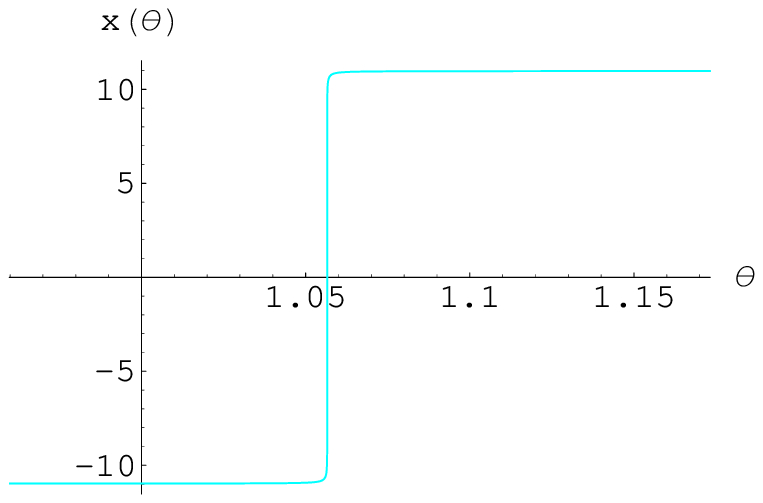}}}
\begin{picture}(0,0)
\end{picture}
\caption{\small The U-shaped solution of the D5-brane
for $R=1$, $r_0=0.32$, $q=3.5$ and $\nu=0.2$.
}
\label{split-rx-fine}
\end{figure}

In both cases, the two cusps are connected by
a confining flux tube of the gauge theory. We estimate the tension of this
flux tube for a tuned configuration shown in the Fig. \ref{split-rx-fine}
as an example for the U shaped solution. 

This tuned U-shaped D5 brane 
is very similar to the quark and anti-quark meson state configuration
which is obtained by the fundamental string action. 
However the present case is for
the D5 brane tube, so the tension of this tube is not equal to the one
of the meson (\ref{f-tension}).
From the Fig. \ref{split-rx-fine}
the tube sit almost
at a constant $\theta\simeq\theta_{c}$ where $\dot p_{\theta}\simeq 0$.
This behavior is understood from
the fact that $\theta_{c}$, 
which is given above by the solution of (\ref{nu}),
is the extremum of the potential $V_0(\theta)$.

\vspace{.3cm}
Then we can estimate
the tension ($\tau_v$) of the flux tube between the two cusp points 
as follows. From the above solutions for $r(x)$ and $x(\theta)$, 
we can approximate as $\theta=\theta_c$ and $x'(\theta_c)>>1$ in the flux region
(see the right hand side of the Fig. \ref{split-rx-fine}).
Then the D5 brane energy (\ref{u}) can be approximated as follows
\bea \label{u-flux}
U_{\rm flux} &=& {N\over 3\pi^2\alpha'}\sin^3\theta_c
\int dx~e^{\Phi/2}
\sqrt{r_x^2+(r/R)^{4}}~\\
&=&{2N\over 3\pi}\sin^3\theta_c~U_F
\eea
where $U_F$ is the energy of the fundamental string given by (\ref{FstringU}). 
This implies the linear
relation of the energy of the flux and its
length $L_{v-v}$. According to the method given for the meson case,
we obtain 
\be \label{tension}
\tau_{v}=\frac{2N}{3\pi} \sin^{3}\theta_{c} ~\tau_M 
\ee
where $\tau_M =T_F\sqrt{q\over R^4}$ denotes the tension 
between the quark and anti-quark which is given above by (\ref{f-tension}).
The dependence of the flux tube tension
on $\nu$ is seen from the factor
$\sin^{3}\theta_{c}$. 
From equation\req{nu}, we find $\sin^{3}\theta_{c}\simeq 
3\pi\nu/2$
for small $\nu$. This implies that $\tau_{v}(\nu=1/N)$ 
reduces to the same tension with the
one of meson state formed by quark and anti-quark \cite{GY}.
This result has a natural
gauge theory interpretation. When one quark is pulled out from the
$SU(N)$ baryon (a color-singlet), the remaining $N-1$ must be in the
anti-fundamental representation of the gauge group. The flux tube
extending between this bundle and a single quark should then have
the same properties as the standard QCD string which connects a quark
and an antiquark. 
On the other hand,
for $\nu N\equiv k>1$, the k-quarks are bounded by the confinning force
since $\tau_v<k \tau_M$.

\vspace{.3cm}
Here we comment on a
similar formula of the flux tension, which has been obtained in another
confining 4d YM theory represented by AdS blackhole background \cite{cgst,ima-05}.
It is written as
\be \label{tension4}
\sigma_B=\frac{8\pi}{27}N(\g{4}^{2}N)T^{2}
      \nu(1-\nu)~.
\ee
This has a similar $\nu$ dependence to our $\tau_v$ in (\ref{tension}), 
but $\sigma_B$ differs from $\tau_v$ in two points.
(i) While $\sigma_B/N$ increases with the 'tHooft coupling $\lambda$ linearly, 
our $\tau_v/N$ is independent of $\lambda$. (ii) The scale of 
our $\tau_v$ is determined by $\langle F_{\mu\nu}^2\rangle$ which is independent
of the compact space scale, 
but $\sigma_B$ is determined by the compact $U(1)$ scale $T$.

\vspace{.5cm}
It would be useful to write the baryon energy or the mass formula for the
split baryon by separating it to three parts as follows,
\beq
E_B^{sp}=N\left(\nu E_F(r_{c_1})+(1-\nu) E_F(r_{c_2})\right)
+\tau_5 L_{v-v}+E_J^{(\nu)}+E_J^{(1-\nu)}
\eeq
where $L_{v-v}$ is the flux length and
$E_F(r_{c_i})$ are given by the same form with (\ref{F-energy}) 
for the two different cusp points $r_{c_i}$, $i=1$ and $2$. 
The last two terms describes the D5 brane parts between the cusps and the 
D5 flux, we call them as "junction" here. They might be given as
\beq
E_J^{(\nu)}\simeq {N\over 3\pi^2\alpha'}\int_0^{\theta_c} d\theta~e^{\Phi/2}
({r^2+r^{\prime 2}})^{1/2}\,
\sqrt{V_{\nu}(\theta)}~,
\eeq
\beq
E_J^{(1-\nu)}\simeq {N\over 3\pi^2\alpha'}\int^{\pi}_{\theta_c} d\theta~e^{\Phi/2}
({r^2+r^{\prime 2}})^{1/2}\,
\sqrt{V_{\nu}(\theta)}~.
\eeq
In the symmetric case of $\nu =1/2$, we can set as $r_{c_1}=r_{c_2}=r_c$, then
the above formula are simplified as
\beq
E=N E_F(r_{c})
+\tau_5 L+2E_J^{(1/2)}
\eeq
It is an interesting problem to compare the junction energy 
$2E_J^{(1/2)}$ with the point
vertex case at $L=0$ and at the same $r_c$. From the above formula, we obtain
at $r_c$
\be
2E_J^{(1/2)}=2 {N\over 3\pi^2\alpha'}(4q)^{1/4}j^{1/2}(\pi/2)~,
\ee
where
\be
 j^{1/2}(\pi/2)=\int_0^{\pi/2}\sqrt{V_{0.5}(\theta)}=4.214~,
\ee
Then the resultant junction energy is compared 
with $E_J^{(S)}$ given by (\ref{junction-e}), and we
find
\be
2E_J^{(1/2)}={2j^{1/2}(\pi/2)\over j(0)}E_J^{(S)}=0.79 E_J^{(S)}~.
\ee
This implies that the split vertex is energetically favourable rather than
the point vertex when the flux length $L_{v-v}$ is negligible.

\vspace{.3cm}
It would be an interesting problem to consider a 
D5 brane configuration which splits to more number of fractions of string fluxes.
An approach in this direction has been given by Imamura \cite{ima-05,ima-04}, and
he has proposed the junction as the vertex part of the splitted 
flux of the fundamental strings. 
Imamura has estimated this part by a numerical calculation under an appropriate
assumptions. 

In the present model, however, it means to study
the three or more number of split D5 vertex. It would be a difficult 
task to obtain
such a solution as a smooth numerical solution.
In this sense, this problem
is the open one here.

\section{Finite Temperature and k-quarks Baryon}
Here, we consider the baryon configurations in the non-confining Yang-Mills theory.
Such a model is given by
the AdS blackhole solution, which represents the high temperature
gauge theory. In our theory with dilaton, the corresponding
background solution is given as
\cite{GSUY}
\beq
ds^2_{10}= e^{\Phi/2}
\left(
\frac{r^2}{R^2}\left[-f^2dt^2+(dx^i)^2\right] +
\frac{R^2}{r^2f^2} dr^2+R^2 d\Omega_5^2 \right) \ .
\label{finite-T}
\eeq

\beq 
f=\sqrt{1-({r_T\over r})^4}, \quad e^\Phi= 
1+\frac{q}{r_T^4}\log({1\over f^2}) \ , \quad \chi=-e^{-\Phi}+\chi_0 \ ,
\label{finite -T-dilaton} 
\eeq
The temperature ($T$) is denoted by $T=r_T/(\pi R^2)$. The world volume 
action of D5 brane is rewritten by eliminating the $U(1)$ flux in terms
of its equation of motion as above, then we get its energy as
\be \label{u-T}
U = {N\over 3\pi^2\alpha'}\int d\theta~e^{\Phi/2} f
\sqrt{r^2+r^{\prime 2}/f^2 +(r/R)^{4}x^{\prime 2}}\,
\sqrt{V_{\nu}(\theta)}~.
\ee
where $V$ is the same form with (\ref{PotentialV}).

\begin{figure}[htb]
\centerline{
{\epsfxsize=14cm
\epsfbox{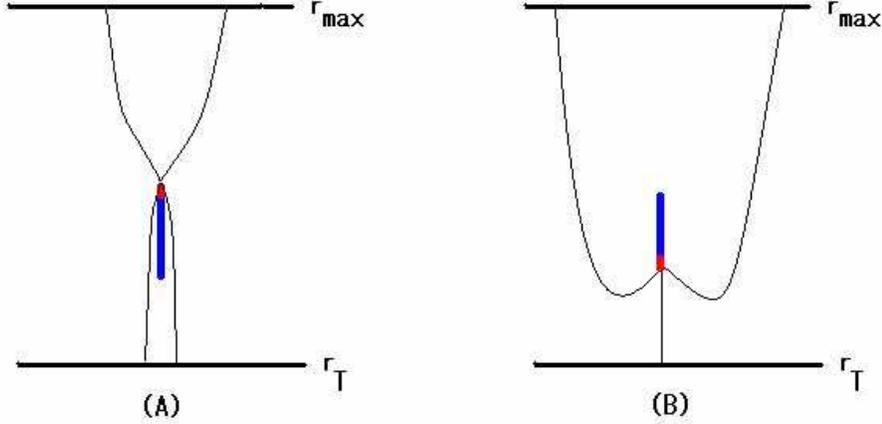}
}}
\caption{\small k-quarks baryons at finite temperature.} 
\label{colored-baryon3}
\end{figure}

For simplicity,
we concentrate on the point vertex configuration. 
Then, we set $x'=0$ as above, and
the equation of motion of $r(\theta)$ is obtained as follows
\beq\label{r-eq-T}
\partial_{\theta}\left({r^{\prime}\over\sqrt{r^2f^2+(r^{\prime})^2}}\,
\sqrt{V_{\nu}(\theta)}\right)
-
{g\over\sqrt{r^2f^2+(r^{\prime})^2}}\,
\sqrt{V_{\nu}(\theta)}=0~.
\eeq
\beq\label{g-func}
g={1\over 2e^{\Phi}}\partial_r\left(e^{\Phi}r^2f^2\right)
\eeq
In this case also, we find the two types of solutions (A) and (B) which have
been given in the section \ref{point3dsec} for 
$T=0$ confinement phase. 
However, in the present
finite $T$ case, the theory is in the quark deconfinement phase. So
free F-strings could exist and these
F-strings touch on the horizon $r_T$. 
As a result, $N-k$ quarks disappear, and we find $k$-quarks baryon. 
This implies that we could find the color non-singlet baryons as
depicted in the Fig.\ref{colored-baryon3}.
This baryon is therefore constructed from $k(<N_c)$ quarks and the 
color singlet D5 vertex. 

\vspace{.3cm}
For fixed $k$, the solution (A) is obtained at small baryon mass or small
$r_0$ where $g$ is positive. When the mass becomes large, then $g$ changes to 
negative value and we find the solution (B). From Eq.(\ref{g-func}), we can
estimate the temperature ($T_{c_1}$) where the solution changes from (A) to (B) as,
\beq
 T<{\gamma\over \pi R^2}q^{1/4}\equiv T_{c_1}\, , \quad \gamma=0.579\, ,
\eeq 
where $\gamma$ is obtained through a numerical analysis.
In any case, we would find $k(<N_c)$ quarks baryon at finite temperature 
before it resolves to independent quarks completely at enough high temperature.
This point is different from the meson, which
is broken from the meson to the quark and antiquark at high temperature and there
is no midlle state as in the case of baryons.

\vspace{.3cm}
Next, we consider the no force condition at the cusp of the k-quarks baryon
for the configurations given in the Fig.\ref{colored-baryon3}. 
The tension of D5 brane at $r_c$ is given as
\be \label{varcalc-T}
\frac{\partial U}{\partial r_{c}}
 =NT_F e^{\Phi/2}\frac{r_{c}^{\prime}}{\sqrt{r_{c}^{\prime 2}+r_{c}^2f(r_c)^2}}~,
\ee
and for F-strings as
\be \label{FstringdU-T}
\frac{\partial U_F}{\partial r_{c}}
 =T_F e^{\Phi/2}\frac{r_{x}^{(c)}}{\sqrt{{r_{x}^{(c)}}^{2}+(r_{c}/R)^4f(r_c)^2}}~,
\ee
Here, for the (N-k) strings ending on the horizon in the Fig.(A),
we consider the limit of the vertical lines as the one in the Fig.(B).
Then the no force condition is obtained as
\beq\label{no-force-T}
 N\frac{r_{c}^{\prime}}{\sqrt{r_{c}^{\prime 2}+r_{c}^2f(r_c)^2}}+(N-k)
 =k\frac{r_{x}^{(c)}}{\sqrt{{r_{x}^{(c)}}^{2}+(r_{c}/R)^4f(r_c)^2}}
\eeq
Notice that $r_x^{(c)}$ and $r_c^{\prime}$ are positive for the solution (A)
and negative for (B) respectively. Then, from $0<k<N$, the 
no force condition Eq.(\ref{no-force-T}) is rewritten as
\beq\label{no-force-2}
  r_x^{(c)} > {r_c\over R^2} r_c^{\prime} ~.
\eeq
When this is compared with the no force condition (\ref{no-force}) for
the confinement phase, we can understand the above condition (\ref{no-force-2})
as a reasonable one. The force of the remaining k-quarks must cover the 
lacked part of (N-k) quarks, so it must become larger.

\vspace{.3cm}
Although the other complicated configurations are possible, 
these baryon configurations would be observed just after the deconfinement
transition occured at high temperature, $T=T_c^{\rm (dec)}$.
The temperature increases above $T_c^{\rm (dec)}$ and 
nears $q^{1/4}$, then the type (B) k-strings baryons will disappear firstly
and the type (A) configurations are remained.
When the temperature increases furthermore, the cusp point arrives at $r_{\rm max}$
at the temperature $T_{\rm melt}$. Since $r_c<r_{\rm max}$,
all the k quarks (including N quarks) baryons should be
collapsed
to the free quarks in the medium of quark gluon plasma for $T>T_{\rm melt}$. 
In other words, k-quarks  
baryons are observed in a range of the temperature,
$$T_c^{\rm (dec)}<T< T_{\rm melt} $$
We need some qualitative and phenomenological analyses to estimate
this temperature range. This point is very interesting, but it is opened
here.

\vspace{.3cm}
On the other hand,
similar k-quarks baryon configurations have been proposed with the
restriction, $k\geq 5N/8$, for $\cal{N}=$ $4$ SYM theory
\cite{baryonsugra,imafirst}. In this theroy, the k-quarks baryon
is possible since the quarks are not confined. But, the
authors in \cite{baryonsugra,imafirst} have not considered the inner
structure of the D5 brane with dissolved F-strings.
In our high temperature model, insteadly, the k-quarks
baryon is possible for
any number of $k$ with $k<N$. This difference
between the condition given in \cite{baryonsugra,imafirst} and ours could
be reduced to the fact that the D5 brane structure is considered or not in
deriving the no-force condition.

Actually, the no-force condition (\ref{no-force-T}) is rewritten as
\beq\label{no-force-T2}
 N(1+Q_5)=k(1+Q_F)\, ,
\eeq
\beq
Q_5=\frac{r_{c}^{\prime}}{\sqrt{r_{c}^{\prime 2}+r_{c}^2f(r_c)^2}}\, , \quad
Q_F=\frac{r_{x}^{(c)}}{\sqrt{{r_{x}^{(c)}}^{2}+(r_{c}/R)^4f(r_c)^2}}\, .
\eeq
Since $\vert Q_F \vert\leq 1$, we obtain
\beq
 \vert Q_F \vert=\left\vert{N(1+Q_5)-k\over k}\right\vert\leq 1\, .
\eeq
From this, we obtain the lower bound of k as
\beq
 {N\over 2}(1+Q_5)\leq k\, .
\eeq
In the case of structureless D5 brane, we obtain $Q_5=1/4$ 
\cite{baryonsugra,imafirst}, then we find $5N/8\leq k$. However, in our
model, $Q_5$ could approach to -1 since $r_c$ could approach to the horizon
$r_T$. This implies the lower bound of k is zero. 

But we must notice that
we need infinite energy to realize the limit of k=0 since the vertex energy
$U$ approaches to infinity in this limit. When the energy of the baryon
increases, its energy is used mainly to extend the length of the F-strings,
namely the value of $L_{q-v}$. However, at finite temperature and in the
deconfinement phase, $L_{q-v}$ has
its maximum value due to the color screening. In this sense, the lower bound of k
would be small but finite for $T> T^{\rm dec}$. So we expect k=0 in the
limit of the $T\to T^{\rm dec}$.
In our model, however, the deconfinement temperature is $T=0$,
so we could find k=0 in the limit of $T\to 0$. In other words, small k state
is seen just above $T^{\rm dec}$, and the lower bound of 
k increases with $T$. So at enough high
temperature, which would be below $T^{\rm melt}$ mentioned above, 
on the other hand, we can not see any $k(<N)$-quarks baryon and
only the k=N baryon is allowed there. So the lower bound of k is dependent on
the temperature.
It is important to assure the details of this statement in a more realistic model
with a finite $T^{\rm dec}$. This point is opened here.

\vspace{.3cm}
The estimations of the mass of these states would be
reported in the next work.

\section{Summary and Discussion}\label{comparison}
The baryon configuration
is studied based on the type IIB string theory 
by embedding D5-brane as a probe in the background corresponding to
two kinds of large $N$ Yang-Mills theories, 
the confining and the deconfining gauge theories. 
The D5 brane is needed as the baryon vertex of the quarks to make
a color singlet, and the structure of the vertex is studied by solving the
embedding equations for the 
fifth coordinate ($r$) and a direction ($x$) of our three dimensional space.

As for the $x$ direction, two typical configurations, the 
point-like and the splitted vertex, are given.
In the latter case, the quarks are separated into two clumps and 
a color flux tube is running between them. 
As found in other confining models, the energy of
such a configuration is proportional to the separation between the 
two quark bundles. And its tension ($\tau_v$), the energy per unit length, is given
by a similar formula given before for the confining theory.
Estimating $\tau_v$, we find a natural dependence on the color charges 
of the individual clumps and that this flux is interpreted as a bound state of 
a number of fundamental strings with a finite
binding energy. The vertex
energy is then given by the length of this flux times $\tau_v$.

As for the $r$-direction, we first show for the point
vertex case. We find that the configurations are 
charachterized by the relation of the
position of the cusp(s) ($r_c$) and the 
extremum point of the D5 brane volume $r_0$, as
(A) $r_c>r_0$ and (B) $r_c<r_0$. These configurations represent the same baryon
at different energy (or mass) state. 
The configuration (A) corresponds to the low energy state and
(B) does for high energy one. So the configuration of the fundamental strings
in $r$-$x$ plane changes when the baryon energy increases. 
Considering of both configurations, the relation of 
the baryon energy ($E$), which is the sum of D5 brane and fundamental strings, 
and the distance ($L_{q-v}$) between 
a fundamental quark in the baryon and the vertex is examined. And we find
a linear relation of $E$ and $L_{q-v}$ at large $L_{q-v}$. The tension
in the case is given by the one of the meson times the kuark number $N$ since
the vertex energy is almost constant at large $L_{q-v}$.
In obtaining this relation, the two vertex configurations (A) and (B) mentioned
above appears. The configuration (A) is dominant at small $L_{q-v}$, where
the linear relation of $E$ and $L_{q-v}$ is not still seen. At an appropriate
$L_{q-v}$, the vertex configuration changes from (A) to (B) and the linear
relation appears. 

In the case of the split vertex, the configuration should be charachterized
in the $x$-$r$ plane as the U shaped and the cap ($\bigcap$) shaped. The 
U shaped one has two type (A) cusps, and the cap shaped one has two type (B)
cusps. The problem to add the fundamental strings
in this case could be solved by applying the results
obtained in the point vertex case to each cusps of the split vertex.
Then we would observe two kinds of the energy scale for the heavy baryon,
$\tau_M$ and $\tau_v$, for the split vertex baryon.

\vspace{.3cm}
As an example of deconfining Yang Mills theory, finite temperature theory is examined.
At enough high temperature, the baryon collapses to quarks completely. Before
arriving this temperature, we find that there is an intermediate temperature,
where quarks are already not confined but they forms a "baryon" state,
which is composed of k-quarks with $k<N$ and color singlet vertex.
So this
is a color non-singlet baryon. This baryon state would be made just after the
phase transition of quark confinement and deconfinement. It is very
interesting problem whether we could find this kind of baryon at high
temperature.

As the next step, we should introduce the probe brane of the quarks and
solve its embedding problem consistently with our baryon configurations
obtaine here. Another important problem is to quantize the baryon system
to obtain their mass spectrum. Some approaches in this direnction are seen
\cite{Brodsky:1997dh,Kirsch:2006he,BST}

\section*{Acknowledgements}
K. Ghoroku thanks to I. Kirsch for useful discussions and comments. 
K.G and M.Ishihara would like to thank F. Toyoda, M. Imachi and S. Otsuki
for fruitful discussions.

\vspace{.5cm}

\end{document}